\documentclass[aps,prl,twocolumn,superscriptaddress,nofootinbib,floatfix,nolongbibliography]{revtex4-2}
\usepackage{amsmath,amssymb,graphicx,booktabs}
\usepackage[hidelinks]{hyperref}
\usepackage{xurl}
\graphicspath{{./}{figs/}}
\newcommand{\M}{\mathcal{M}}
\newcommand{\lamP}{\lambda_{\mathrm{P}}}
\newcommand{\Ws}{\widetilde W_{s}}
\newcommand{\Wc}{\widetilde W_{\mathrm{coll}}}
\newcommand{\ft}{\tilde f}
\newcommand{\bk}{\mathbf k}
\newcommand{\bq}{\mathbf q}
\newcommand{\bx}{\mathbf x}

\begin{document}

\title{More particles or the pair? An exact test of how to improve mode-coupling theory}

\author{Hu Cang}
\email{cangh@uci.edu}
\affiliation{Department of Developmental and Cell Biology, University of California, Irvine, Irvine, California 92697, USA}

\begin{abstract}
How a dense liquid arrests into a glass remains a central problem in physics. Mode-coupling theory (MCT), the leading microscopic theory of the glass transition, fails to reproduce mean-field caging. Two routes are proposed to repair it: tracking more-particle correlations, or enforcing repulsion within each colliding pair. Testing both on an exact benchmark liquid, we show that MCT lets touching spheres pass through each other at all orders: adding particles dilutes memory and worsens the transition, whereas restoring pair impenetrability recovers mean-field arrest and the cage size.
\end{abstract}

\maketitle

\textit{Introduction.} As a liquid is cooled or compressed toward the glass transition, its relaxation slows by many orders of magnitude without any major change in static structure~\cite{Debenedetti2001,Berthier2011}. Microscopically, this slowdown is driven by caging: each particle is trapped in a local cage of neighbors until the cage relaxes~\cite{Gotze2009,Weeks2000,Cang2026Contacts,Cang2026Shared}. Mode-coupling theory (MCT) is the leading first-principles framework that predicts this arrest from equilibrium liquid structure alone, closing the dynamical equations at pair density correlations (order $N=2$, where $N$ is the maximum number of correlated density modes tracked)~\cite{Bengtzelius1984,Leutheusser1984,Gotze2009,Cang2003PRL,Cang2005}. For hard spheres of packing fraction $\varphi$ (occupied volume fraction), standard $N=2$ MCT predicts permanent caging at a critical packing fraction $\varphi_c=0.516$~\cite{Franosch1997,Gotze2009}, below the observed glass crossover $\varphi_\times\approx 0.58$--$0.59$~\cite{Pusey1987,Brambilla2009}. Generalized MCT (GMCT) extends MCT to higher multi-particle orders $N>2$~\cite{Szamel2003,Wu2005,Mayer2006,Janssen2015,Janssen2016,Janssen2018}; using the standard Percus--Yevick (PY) static structure factor for hard spheres, $\varphi_c$ rises from $0.516$ ($N=2$) to $0.565$ ($N=6$, memory closure~\cite{Luo2020a,Luo2020b}), and from $0.561$ ($N=6$) to $0.567$ ($N=7$, this work) in the correlator closure, a rise Luo and Janssen call ``seemingly convergent (approximately logarithmic)'' while anticipating that it may continue ``perhaps indefinitely''~\cite{Luo2020a}.

\begin{figure}[t]
\centering\includegraphics[width=\columnwidth]{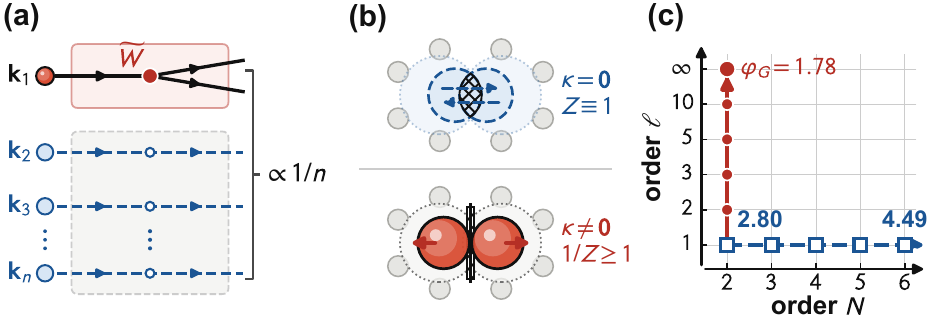}
\caption{Two competing routes beyond the $N=2$ mode-coupling kernel. (a)~Multi-particle order $N$: at the $n$-particle level ($n\le N$), one leg collides (vertex $\widetilde W$) against $n$ additive bare diffusive rates $\sum_{i=1}^n D_0 k_i^2$, bounding memory by $C/n$ ($C$: single-pair coupling constant). (b)~Pair hard core: standard mode-coupling propagation (relative wavevector transfer $\boldsymbol\kappa=\mathbf 0$, top) erases the hard core between bonded neighbors (pair non-overlap probability $Z\equiv 1$); keeping intra-pair reflection ($\boldsymbol\kappa\ne\mathbf 0$, bottom; simulated bond correlation $R\approx 2$) weights memory by $1/Z(\mathbf c)\ge 1$, recovering the mean-field transition ($\varphi_G=1.78$). (c)~Critical arrest packing fractions $\varphi_c$ in the MK liquid: adding $N$ keeps $\varphi_c$ moving away ($2.80\to 4.49$); intra-pair exclusion orders $\ell$ converge to $\varphi_G$ [overview in Supplemental Material (SM), Sec.~S6].}
\label{fig:cartoon}
\end{figure}

Whether this shift is true convergence cannot be decided on ordinary hard spheres, where static structural approximations mix with dynamical closures~\cite{Luo2022,Pihlajamaa2023,Pihlajamaa2024,Pihlajamaa2024b,Pihlajamaa2025} and the exact caging threshold is unknown. A foundational limit where caging is known from first principles is mean-field caging, established by Parisi's replica theory and cavity methods~\cite{Parisi2010,Charbonneau2014,Charbonneau2017}: because a particle's neighbors are statistically independent and form no closed contact loops, the cage follows from the pair interaction alone. This description is exact as the spatial dimension $d\to\infty$~\cite{Maimbourg2016,Agoritsas2019}, and its loop-free premise is realized in $d=3$ by the Mari--Kurchan (MK) benchmark liquid, where each pair interacts at an independent random shift so that closed contact loops vanish and the equilibrium structure is exact~\cite{Mari2009,Mari2011,Nishikawa2022,Charbonneau2024}. Any dynamical error is thus strictly the theory's own. Standard $N=2$ MCT fails this test in high $d$~\cite{Ikeda2010,Schmid2010,Maimbourg2016} and on the $d=3$ MK liquid (in our shift-aware closure below), predicting arrest at $\varphi_c=2.80$, far above the Gaussian mean-field transition $\varphi_G=1.78$~\cite{Charbonneau2014,Szamel2017} (SM Sec.~S5\,E).

Two competing routes have been proposed to repair MCT: adding correlations among ever more particles $N$ (GMCT~\cite{Szamel2003,Wu2005,Mayer2006,Janssen2015,Luo2020a}), or retaining direct repulsion within each colliding pair (first explored in a two-particle dynamical cavity model by Szamel~\cite{Szamel2017}, where keeping the intra-pair force recovers the Gaussian replica plateau equation and dropping it yields the Kirkpatrick--Wolynes localization equation~\cite{Kirkpatrick1987}). Building on this cavity insight, we extend the analysis to all multi-particle orders $N$ in GMCT and to simulations, testing both routes head-to-head on the MK liquid to establish four results showing that pair impenetrability, not order $N$, governs mean-field caging.

\begin{figure*}[t]
\includegraphics[width=\textwidth]{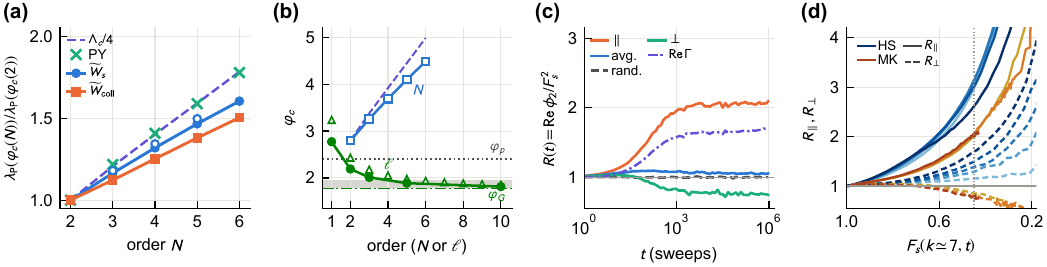}
\caption{Multi-particle order $N$ vs.\ intra-pair exclusion order $\ell$, and the pair correlation that MCT and GMCT miss. (a)~Feedback ratio $\lamP(\varphi_c(N))/\lamP(\varphi_c(2))$ vs.\ $1/n$ dilution law $\Lambda_c(N)/4$ (dashed): PY hard spheres (crosses; memory closure~\cite{Luo2020a}) agree to $1\%$; on MK, the self vertex $\Ws$ falls below by $3.5$--$10\%$ (filled circles, correlator) or $2.2$--$5.9\%$ (open circles, memory) and collective $\Wc$ (squares, correlator) by $7$--$16\%$. (b)~MK critical packing fraction $\varphi_c$ vs.\ order $N$ (squares) and exclusion order $\ell$ [Eq.~\eqref{eq:series}; $k$-resolved, circles, from $2.77$ (continuum $N=2$); Gaussian, triangles, from $3.24$], dilution law (dashed), simulated crossover $\varphi_\times\in[1.80,1.93]$ (shaded~\cite{Mari2011,Charbonneau2014}), static cage percolation $\varphi_p=2.40$ (dotted~\cite{Charbonneau2014}), and mean-field transition $\varphi_G=1.78$ (dash-dotted). (c)~Pair correlator $\phi_2(t)$ at $\varphi=2.0$ ($|\bk_1|=|\bk_2|\simeq 7$, $\bk_2=-\bk_1$) as factorization ratio $R(t)={\rm Re}\,\phi_2/F_s^2$ ($R=1$: independent motion) for random (dashed), orientation-averaged contact (avg., blue), transverse ($\bk_1\perp\hat{\mathbf r}_{ij}$, green), and longitudinal ($\bk_1\parallel\hat{\mathbf r}_{ij}$, orange, $R\approx 2$) pairs; dash-dotted: connected two-body correlation ${\rm Re}\,\Gamma$ (${\rm Re}\,\Gamma=1$: independent recoil). (d)~Longitudinal ($R_\parallel$, solid) and transverse ($R_\perp$, dashed) ratios vs.\ $F_s(k\simeq7,t)$ in 3D hard spheres ($\varphi=0.50$--$0.60$, blue) and MK ($\varphi=1.5, 1.8, 2.0$, oranges; dotted: $F_s(7)=0.45$).}
\label{fig:runoff}
\end{figure*}

First, we prove by machine-checked theorems in Lean~4 (SM Sec.~S1\,B--D) that caging memory at the $n$-particle level is capped at $C/n$ [Fig.~\ref{fig:cartoon}(a)], so $\varphi_c(N)$ cannot converge on any fixed grid; solving GMCT for $N=3$--$6$ shows that $\varphi_c$ moves farther from $\varphi_G=1.78$ ($2.80\to 4.49$) [Figs.~\ref{fig:cartoon}(c), \ref{fig:runoff}(b)]. Second, standard mode-coupling propagation erases the reflecting wall between colliding neighbors and lets touching spheres pass through each other like ghosts [Fig.~\ref{fig:cartoon}(b)]; an all-order wavevector invariant shows that orders $N>2$ in GMCT never cure this flaw.

Third, enforcing hard-core impenetrability within a single colliding pair at lowest order ($N=2$) amplifies the caging kernel by $1/Z(\mathbf c)\ge 1$, where $0<Z(\mathbf c)\le 1$ is the probability that two neighbors rattling in their cages do not overlap ($Z\equiv 1$ in standard MCT), lowering $\varphi_c$ from $2.80$ to $\varphi_G=1.78$ [Figs.~\ref{fig:cartoon}(c), \ref{fig:runoff}(b)]; its Gaussian cage reduction recovers the replica and cavity plateau equation~\cite{Parisi2010,Mari2011,Szamel2017} and predicts the simulated mean cage size across eight arrested densities within $5\%$. Fourth, direct simulations of the MK liquid and ordinary hard spheres show that along the contact bond where the caging force acts, the pair correlator $\phi_2(t)$ is two to four times the independent-motion product $F_s(k,t)^2$ (MCT's factorization on MK) [Fig.~\ref{fig:runoff}(c,d)], the dynamical signature of the missing pair hard core on MK (exact statics), whereas a pair-based reconstruction captures the displacement covariances of touching triplets (SM Sec.~S4\,F).

\textit{Every added order dilutes the caging memory.} Why does increasing the multi-particle (closure) order $N$ push the predicted arrest to higher density? At the $n$-particle level ($n\le N$), the $n$-particle correlator tracks $n$ density modes $K=(\bk_1,\dots,\bk_n)$ (wavenumbers $k_m=|\bk_m|$ in inverse sphere diameters) in parallel. Because diffusion of any of the $n$ particles relaxes the $n$-particle correlation, their bare relaxation rates add in the denominator; in contrast, as shown in Fig.~\ref{fig:cartoon}(a), a pair force acts on only one pair at a time, so the numerator has only one active collision vertex $\widetilde W$ while the $n-1$ spectator modes $K\setminus k_m$ coast freely. On the discrete grid, this competition governs the stationary GMCT equations~\cite{Szamel2003,Wu2005,Franosch2002,Janssen2015,Luo2020a,Biezemans2020} for the $n$-particle plateau $f_n(K)\in[0,1)$ ($f_n=0$: liquid; $f_n>0$: glass) and kernel $\M_n$ (SM Sec.~S1\,A):
\begin{gather}
\frac{f_n(K)}{1-f_n(K)}=\M_n(K)[f_{n+1}],\label{eq:hier}\\
\M_n(K)[g]\!=\!\!\sum_{m=1}^{n}\!\frac{\Omega_1^2(k_m)}{\Omega_n^4(K)}\!\sum_{q,p}\!\widetilde W(k_m;\!q,\!p)g(q,\!p,\!K{\setminus}k_m)\!,\label{eq:mem}
\end{gather}
where $g$ is the $(n+1)$-particle input [$g=f_{n+1}$ in Eq.~\eqref{eq:hier}], $\Omega_n^2(K)=\sum_{m=1}^n\Omega_1^2(k_m)$ sums the single-mode bare rates $\Omega_1^2(k)=D_0k^2/S(k)$ ($D_0$: bare diffusion constant; $S(k)$: static structure factor, $S(k)\equiv 1$ in MK), and $\widetilde W(k_m;q,p)\ge0$ is the static pair-collision vertex. At closure order $N$ ($N=2$: standard MCT), GMCT closes Eq.~\eqref{eq:hier} by factorizing either the $N$-particle correlator $f_N$ or the memory kernel $\M_{N-1}$ into products of lower-order functions (correlator and memory closures; SM Sec.~S1\,A). By $\sum_{m=1}^n\Omega_1^2(k_m)/\Omega_n^4(K)=1/\Omega_n^2(K)\le 1/(n\omega_{\min})$ ($\omega_{\min}=\min_k\Omega_1^2(k)$), caging memory at level $n$ is diluted as $1/n$ [SM Sec.~S1\,C, Eq.~(S9)].

With the single-pair coupling constant $C=w_{\max}/\omega_{\min}$ (maximal vertex row sum $w_{\max}$, slowest rate $\omega_{\min}$), we prove three results on any fixed wavenumber grid (SM Sec.~S1\,B--C): (i)~the $n$-particle memory kernel obeys $\|\M_n[g]\|\le (C/n)\|g\|$; (ii)~arrest at order $N$ requires $C\ge N-1$ (correlator) or $2C\ge N-2$ (memory), so $\varphi_c(N)$ cannot converge to any density where $C$ stays finite ($\varphi_c(N)\to\infty$ on MK where $C\propto\varphi$); and (iii)~the $N\to\infty$ limit has no glass plateau at finite $C$, extending schematic results~\cite{Mayer2006,Janssen2016}. The finite-order bounds and factorial estimate [SM Eq.~(S8), implying (iii)] are machine-checked in Lean~4~\cite{Moura2021} (SM Sec.~S1\,D) and rule out fixed-grid convergence with $N$ (binding for $N-1\gtrsim C\approx2.5\times10^3\varphi$ on the MK grid; SM Table~S3).

For accessible orders $N\le 7$, $1/n$ dilution acts mainly on the leading eigenmode of the frozen $N=2$ memory-feedback operator, whose Perron eigenvalue $\lamP(\varphi)$ measures two-body caging feedback [SM Eq.~(S6), Sec.~S1\,E]. Because higher orders $N>2$ mainly pit this single $N=2$ cage mode against $1/n$ dilution, arrest requires a stronger feedback $\lamP(\varphi_c(N))/\lamP(\varphi_c(2))\simeq\Lambda_c(N)/4$, where $\Lambda_c(N)/4>1$ is the dilution factor of single-mode GMCT~\cite{Mayer2006,Janssen2016} ($\Lambda_c(2)=4$). As shown in Fig.~\ref{fig:runoff}(a), published PY memory-closure thresholds for $N=3$--$6$~\cite{Luo2020a} (crosses) follow this parameter-free dilution line (dashed) within $0.85\%$ (mode overlap $99.8\%$ at $k\approx 7$ for $N\le 5$); our correlator-closure series to $N=7$ ($\varphi_c=0.5159\to 0.5671$) rises monotonically with a growing offset ($-1.3\%$ to $-7.2\%$; SM Sec.~S1\,E, Table~S5). The apparent convergence in PY hard spheres thus reflects single-mode $1/n$ dilution rather than multi-particle caging. Inverting the same law beyond the solved orders suggests continued drift ($\varphi_c\approx 0.58$ at $N\approx9$--$10$ and $0.59$ at $N\approx11$--$13$; SM Sec.~S1\,E, Table~S5), so matching one value selects an order rather than testing the closure.

\textit{A benchmark with mean-field caging.} In ordinary hard spheres, static approximations mix with dynamical closures~\cite{Mangeat2016}, masking whether higher orders $N$ improve caging (SM Sec.~S5\,F). The 3D MK model~\cite{Mari2009,Mari2011} cleanly separates statics from dynamics: each pair of unit spheres at number density $\rho$ ($\varphi=\pi\rho/6$) interacts via a quenched random shift that eliminates closed contact loops, making the contact network tree-like and static structure exact ($S(k)=1$; SM Sec.~S2\,A). In $d=3$, caging is anchored by three benchmarks in Fig.~\ref{fig:runoff}(b)~\cite{Parisi2010,Mari2011,Charbonneau2014}: the mean-field (Gaussian replica) transition $\varphi_G=1.78$ (dash-dotted; SM Sec.~S5\,E), simulated crossover $\varphi_\times\in[1.80,1.93]$ (shaded), and cage percolation $\varphi_p=2.40$ (dotted).

Because bulk density modes have $S(k)=1$ and yield a vanishing kernel in standard MCT, we construct shift-aware $n$-particle tree modes in the shifted pair frame and project the dynamics onto this basis using exact ensemble symmetries (SM Sec.~S2\,B--D). This yields the MK GMCT equations in Eqs.~\eqref{eq:hier}--\eqref{eq:mem} with $\Omega_1^2(k)=D_0k^2$ and the $N=2$ kernel:
\begin{equation}
\M_1(k,t)=\!\int\!\frac{\rho\,d^3q}{(2\pi)^3}\frac{(\hat\bk{\cdot}\bq)^2}{k^2}\ft(q)^2 F_s(|\bk{-}\bq|,\!t)F_s(q,\!t),\label{eq:kernel}
\end{equation}
where $\hat\bk=\bk/k$, $\ft(q)$ is the Fourier transform of the hard-core Mayer function $f(r)=-\Theta(1-r)$ (SM Sec.~S2\,A), and $F_s(q,t)=\langle e^{i\bq\cdot\Delta\bx_i(t)}\rangle$ is the single-particle density correlator [with glass plateau $f_s(q)=F_s(q,\infty)$]. Because every static input in Eq.~\eqref{eq:kernel} is exact, any error comes purely from the dynamical closure. Moreover, because the collision vertex in Eq.~\eqref{eq:mem} scales linearly with $\rho$ (for both the self vertex $\Ws$ in Eq.~\eqref{eq:kernel} and collective vertex $\Wc$; SM Sec.~S2\,D--E), $C,\lamP\propto\varphi$ and single-mode dilution predicts $\varphi_c(N)/\varphi_c(2)\simeq\Lambda_c(N)/4$.

\textit{Higher orders move away from mean-field caging.} Because each added level $n$ dilutes caging memory as $1/n$, the weakened kernel requires a higher collision rate (and higher $\varphi$) to lock particles into a glass. Solving our exact-statics MK GMCT equations for $N=2$--$6$ confirms that every order arrests above the caging regime and moves farther away [squares in Figs.~\ref{fig:cartoon}(c) and \ref{fig:runoff}(b); SM Sec.~S3]: $\varphi_c(N)$ rises from $2.80$ ($N=2$) to $3.26, 3.69, 4.10, 4.49$ ($N=3$--$6$), more than twice $\varphi_\times$ for $N\ge5$ (grid spacing $h\to 0$: $\varphi_c(2)=2.77$, shifting $<1\%$ for $N\le 4$; collective $\Wc$: $2.41\to 3.62$; SM Tables~S1, S2).

\textit{No order restores the hard core inside a pair.} Why does $N=2$ already overshoot $\varphi_G=1.78$, and why does adding particles fail to fix it? We trace both flaws to the frequency matrix of a colliding pair at total wavevector $\mathbf Q=\bk_1+\bk_2$ [SM Eq.~(S22)]:
\begin{equation}
-D_0(k_1^2+k_2^2)\Big[\delta_{\bk_2\bk_2'}+\frac{\rho\ft(\kappa)}{N_p}\Big]+\frac{D_0\rho}{N_p}(\bk_1-\bk_2)\cdot\boldsymbol\kappa\,\ft(\kappa),\label{eq:K}
\end{equation}
where $\delta_{\bk_2\bk_2'}$ is the Kronecker delta, $N_p$ is the particle count, and $\boldsymbol\kappa=\bk_2'-\bk_2$ is the relative wavevector transfer.

In Eq.~\eqref{eq:K}, the $\boldsymbol\kappa=\mathbf 0$ term describes independent diffusion and supplies the additive bare rates in Eq.~\eqref{eq:mem} that cause $1/n$ memory dilution [Fig.~\ref{fig:cartoon}(a)]. Conversely, the off-diagonal terms ($\boldsymbol\kappa\ne\mathbf 0$) sum over all $\sim N_p$ relative modes (canceling $1/N_p$) to form a finite operator for the pair, where the bracketed term enforces static hard-core exclusion and the last term is the impenetrable reflecting wall at contact~\cite{Cang2026Contacts} [Fig.~\ref{fig:cartoon}(b), bottom; SM Sec.~S2\,F]. As sketched in Fig.~\ref{fig:cartoon}(b) (top), standard mode-coupling propagation retains only the wavevector-diagonal ($\boldsymbol\kappa=\mathbf 0$) term and drops this off-diagonal operator (the intra-pair force in Szamel's cavity model~\cite{Szamel2017}), erasing the reflecting wall ($Z\equiv 1$) and letting touching neighbors pass through each other like ghosts. Because ghost-passing particles escape their cages too easily, $N=2$ MCT requires a much higher density ($\varphi_c=2.80$ vs.\ $\varphi_G=1.78$).

Crucially, we prove that higher orders $N>2$ of diagonal GMCT never restore this reflecting wall. Although Ref.~\cite{Laudicina2026PRE} noted that GMCT is wavevector-diagonal, one might expect higher orders to regenerate the wall effectively; instead, they only attach spectator branches at vertices while keeping inter-vertex propagation diagonal ($\boldsymbol\kappa=\mathbf 0$), transferring no relative wavevector across bonds (SM Sec.~S2\,G).

\textit{Simulations show the pair correlation that MCT and GMCT miss.} Because shifted neighbors in the tree-like MK liquid share no common neighbors, any pair correlation beyond independent motion comes directly from their mutual hard core. In equilibrium (planted) Monte Carlo simulations~\cite{Mari2011} ($N_p=1000$, yielding $\varphi_\times=1.80\pm0.04$; SM Eq.~(S27)), we test the MCT factorization $\phi_2(t)\approx F_s(k_1,t)^2$ in Eq.~\eqref{eq:kernel} for the pair correlator $\phi_2=\langle e^{i\bk_1\cdot\Delta\bx_i}e^{i\bk_2\cdot\Delta\bx_j}\rangle$ ($|\bk_1|=|\bk_2|\simeq7$, $\bk_2=-\bk_1$) via $R(t)={\rm Re}\,\phi_2(t)/F_s(k_1,t)^2$ ($R=1$: factorized; SM Table~S11). In Fig.~\ref{fig:runoff}(c), random pairs factorize ($R=1.00\pm0.01$, dashed), and contacting pairs (unit bond vector $\hat{\mathbf r}_{ij}$) also appear to factorize within $5\%$ on the $\varphi=2.0$ plateau ($5$--$15\%$ across $\varphi\ge 1.2$) under orientation averaging (blue), where transverse ($\bk_1\perp\hat{\mathbf r}_{ij}$, green, $R_\perp\approx0.78$) and longitudinal motions cancel. Resolved along the contact bond ($\bk_1\parallel\hat{\mathbf r}_{ij}$, orange, ratio $R_\parallel$) where the hard-core force acts, however, $\phi_2(t)$ is twice the factorized product $F_s^2$, reaching $R_\parallel=2.3$--$2.6$ at $F_s=1/e$ ($\varphi=0.8$--$1.8$) and $2.02\pm0.05$ on the $\varphi=2.0$ plateau. Dividing out single-particle recoil defines the connected two-body correlation ${\rm Re}\,\Gamma$ (${\rm Re}\,\Gamma=1$: independent recoil; dash-dotted; SM Sec.~S4\,C~\cite{Cang2026Shared}), reaching $1.64\pm0.02$.

Plotted against single-particle decorrelation $F_s(k\simeq7,t)$ in Fig.~\ref{fig:runoff}(d), this underestimation of $\phi_2(t)$ by $F_s^2$ along the bond is even stronger in 3D polydisperse hard spheres ($\varphi=0.50$--$0.60$, blue; MCT-fit crossover $\approx 0.598$~\cite{Berthier2016}; $R_\perp>1$ from slower contact pairs and shared mobility, SM Sec.~S4\,F) than in MK (oranges): at the MK plateau value $F_s(7)=0.45$ (dotted), contact pairs in 3D hard spheres reach $R_\parallel=2.7$--$3.2$ and ${\rm Re}\,\Gamma=1.9$--$2.2$ ($R_\parallel=3.3$--$4.1$ at relaxation). In real space, the Pearson correlation of bond-projected cage displacements ($0$: uncorrelated, $1$: locked; SM Sec.~S4\,F, Table~S14) reaches $0.51$--$0.63$ along the bond vs.\ $\le0.09$ transverse to it on the $\beta$ plateau (MC, Newtonian).

Does dense-liquid caging require genuine three-particle displacement correlations beyond pairs? Testing touching triplets in 3D hard spheres against a Gaussian--Markov superposition of measured pair correlations (including the outer-pair contact covariance in closed triangles; SM Sec.~S4\,F) shows that across $15$ $(\varphi,t)$ states, the superposition captures the bond-projected displacement covariance within a normalized residual of $0.025$ in $12$ cases (at most $0.046$; late-time excess in joint bond breaking).

\textit{Restoring pair impenetrability at lowest order ($N=2$).} To restore pair impenetrability at $N=2$, we enforce hard-core exclusion directly on the glass plateau within a frozen-cage ansatz (SM Sec.~S5\,A). For two bonded spheres in cages separated by $\mathbf c$ ($c=|\mathbf c|$), standard MCT treats their excursions as independent, allowing core overlap [Fig.~\ref{fig:cartoon}(b), top]; excluding overlap [Fig.~\ref{fig:cartoon}(b), bottom] shrinks the allowed two-cage fraction to the non-overlap probability $Z(\mathbf c)\in(0,1]$. Restricting the pair to the allowed fraction $Z(\mathbf c)$ amplifies each vertex by $1/Z(\mathbf c)$ [giving $1/Z(\mathbf c)^2$], and averaging over the cage-center measure $\rho Z(\mathbf c)d^3c$ leaves a weight $1/Z(\mathbf c)\ge 1$:
\begin{equation}
\M_1(k,\infty)=\frac{\rho}{k^2}\int d^3c\,\frac{|\mathcal{A}(k,\mathbf c)|^2}{Z(\mathbf c)},\label{eq:brc}
\end{equation}
where $\mathcal A(k,\mathbf c)$ is the contact collision amplitude [SM Eq.~(S32)]. Setting $Z(\mathbf c)\equiv1$ in Eq.~\eqref{eq:brc} (allowing overlap) recovers the standard $N=2$ kernel in Eq.~\eqref{eq:kernel} by Parseval's theorem (SM Sec.~S5\,A); standard MCT thus fails mean-field caging by setting $1/Z=1$.

To track how pair impenetrability restores caging, we expand $1/Z(\mathbf c)$ in the cage-overlap probability $1-Z(\mathbf c)$ as a series $Y_\ell(\mathbf c)$ of intra-pair exclusion order $\ell$. For an isotropic Gaussian cage of variance $\sigma^2$ per coordinate (MSD $6\sigma^2$), Eq.~\eqref{eq:brc} as $k\to 0$ yields (SM Sec.~S5\,B--C):
\begin{equation}
Y_\ell(\mathbf c)=\sum_{j=0}^{\ell-1}[1-Z(\mathbf c)]^j,\quad \frac{\rho\sigma^2}{3}\int d^3c\,[\partial_c Z(c)]^2\,Y_\ell(c)=1.\label{eq:series}
\end{equation}

In Figs.~\ref{fig:cartoon}(c) and \ref{fig:runoff}(b), intra-pair exclusion order $\ell$ and multi-particle order $N$ start from the same standard $N=2$ kernel ($\ell=1$, $Y_1=1$) and move in opposite directions: whereas adding $N$ (squares) drives $\varphi_c$ upward, keeping $\ell=1, 2, 3, 5, 10$ terms of $Y_\ell$ (circles and triangles) lowers $\varphi_c(\ell)$ from $2.77$ [continuum limit of the grid value $\varphi_c(2)=2.80$; SM Table~S2] across $\varphi_p=2.40$ to $2.19, 2.01, 1.89, 1.815\to 1.784$ ($k$-resolved, circles; $3.24\to 1.776$ for Gaussian, triangles; $3.33\to 1.789$ for the finite-domain static Bethe functional; SM Table~S16). At full exclusion ($\ell\to\infty$), Eq.~\eqref{eq:series} matches the Gaussian replica~\cite{Parisi2010,Mari2011,Charbonneau2014} and Szamel cavity~\cite{Szamel2017} equations ($\varphi_G=1.78$), whereas $Y_1=1$ ($\ell=1$) gives the Kirkpatrick--Wolynes equation~\cite{Kirkpatrick1987,Szamel2017} ($\varphi_c=3.24$). Simulations across eight arrested densities confirm the Gaussian plateau MSD of Eq.~\eqref{eq:series} within $5\%$ without free parameters, as does the static Bethe closure; imposing Eq.~\eqref{eq:brc} pointwise in $k$ gives nearly the same threshold ($1.784$) but a softer plateau (SM Secs.~S4\,D--E, S5\,D~\cite{Weeks2000,Chaudhuri2007,Biroli2022}). In ordinary 3D hard spheres, by contrast, the contact peak of $g(r)$ masks the exclusion weight: within the same Gaussian closure, removing it shifts $\varphi_c$ by only $1\%$ ($0.518\to0.523$ with PY statics) vs.\ $82\%$ on MK (SM Sec.~S5\,F). The restored-pair (Gaussian replica) threshold $\varphi_d^{\rm MZ}=0.518$ (our evaluation; $0.522$ in Ref.~\cite{Mangeat2016}) thus lies next to PY-MCT's $0.516$~\cite{Franosch1997}, motivating corrections beyond this mean-field description to reach $\varphi_\times\approx0.58$--$0.59$~\cite{Laudicina2024,Laudicina2026PRL,Laudicina2026PRE}.

\textit{Conclusion.} Here we have shown why mode-coupling theory fails mean-field caging and how to repair it: standard propagation lets colliding neighbors pass through each other like ghosts while diluting caging memory as $1/n$ at every level $n\le N$, so tracking more particles only weakens the cage. Conversely, restoring impenetrability within a single pair at $N=2$ recovers the mean-field transition~\cite{Szamel2017} and, in its Gaussian and static reductions, predicts the mean cage size parameter-free. Simulations reveal the bond-aligned pair correlation that factorized propagation misses. Pair impenetrability, rather than higher multi-particle orders, is thus the foundation of mean-field glass caging, the skeleton on which finite-dimensional fluctuation theories~\cite{Laudicina2026PRL} can build.

\begin{acknowledgments}
Supported by NIH Grant R01HG014004. Code/data: \url{https://github.com/hucang0/lean_mct} (SM Sec.~S7).
\end{acknowledgments}

\bibliography{refs}

\end{document}